# Adaptive dual-comb spectroscopy in the green region

Takuro Ideguchi,[1] Antonin Poisson,[1,2] Guy Guelachvili,[2] Theodor W. Hänsch,[1,3] Nathalie Picqué[1,2,3,*]

1. Max-Planck-Institut für Quantenoptik, Hans-Kopfermann-Strasse 1, 85748 Garching, Germany
2. Institut des Sciences Moléculaires d'Orsay, CNRS, Bâtiment 350, Université Paris-Sud, 91405 Orsay, France.
3. Ludwig-Maximilians-Universität München, Fakultät für Physik, Schellingstrasse 4/III, 80799 München, Germany.
*Corresponding author e-mail address: nathalie.picque@mpq.mpg.de

**Abstract**
*Dual-comb spectroscopy is extended to the visible spectral range with a set-up based on two frequency-doubled femtosecond ytterbium-doped fiber lasers. The dense rovibronic spectrum of iodine around 19240 $cm^{-1}$ is recorded within 12 ms at Doppler-limited resolution with a simple scheme that only uses free-running femtosecond lasers.*

Laser frequency combs have revolutionized [1,2] the way we measure the frequency of light and are becoming essential for many new applications that rely on the precise control of light waves. A laser frequency comb has a broad spectrum, generally produced by a mode-locked femtosecond laser, which consists of several hundred thousands perfectly evenly spaced spectral lines. Molecular spectroscopy is one of the novel areas of science and technology where these recent light sources may be taken advantage of. In particular, dual-comb spectroscopy [3-9] offers intriguing opportunities, as this broad spectral bandwidth multiplex technique has demonstrated dramatically reduced acquisition rates, improved resolution and sensitivity when compared to the most commonly used tool for broadband molecular spectroscopy, the Michelson-based Fourier transform interferometer [10]. Dual-comb spectroscopy is a time-domain interferometric technique, in which the pulse train of the interrogating comb excites the absorbing sample at regular time intervals. A second pulse train of different repetition frequency interferometrically samples the transient response or "free induction decay" of the medium, akin to an optical sampling oscilloscope. A Fourier transform reveals the spectrum.

The most convincing demonstrations of dual-comb spectroscopy have been undertaken in the near-infrared range [4,5,7,8], where frequency comb oscillators are conveniently available. However, most of molecular transitions in this region are due to weak overtone bands. Recently, significant progress has been reported [3,6] on the extension of this promising technique towards the mid-infrared domain, where its early proof-of-principle demonstrations [9] had also been carried out. The visible region is complementary to the "molecular fingerprint" mid-infrared range: many molecules undergo strong rovibronic transitions in the visible. Moreover, advanced photonic technologies are readily available in this domain. However, dual-comb spectroscopy has not been implemented yet in the visible and ultra-violet regions.

In this letter, we report on real-time dual-comb spectroscopy in the visible spectral region. A new scheme that makes it possible to record distortion-free dual-comb spectra with frequency-doubled free-running femtosecond ytterbium-doped fiber laser systems is implemented. Phase-lock electronics or a posteriori corrections are not required. We investigate the dense rovibronic spectrum of iodine in the 19240 cm$^{-1}$ region at Doppler-limited resolution.

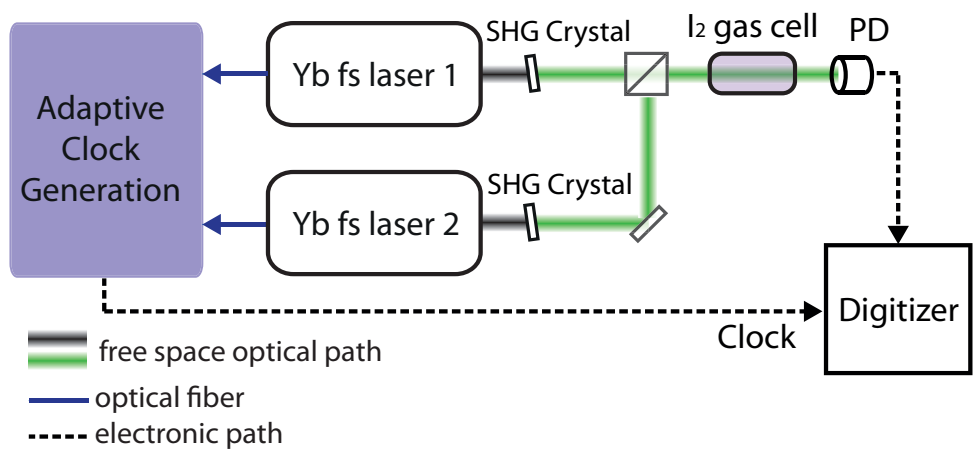

**Figure 1.** Experimental setup. Two free-running femtosecond ytterbium-doped fiber lasers with slightly different repetition frequencies are frequency-doubled with second-harmonic-generation (SHG) crystals and combined. The green beams probe the iodine sample. Their time-domain interference signal is recorded with a fast photodetector PD and digitized using the adaptive clock, described in Fig. 2.

Figure 1 sketches the experimental set-up. Two free-running ytterbium-doped fiber commercial mode-locked oscillators have a central wavelength of 9620 cm$^{-1}$. They emit pulses of about 100 fs with an average power of 60 mW. Their repetition frequencies are about 100 MHz and they differ by 6.7 Hz. For the spectroscopy set-up, the two laser beams are frequency-doubled with 2mm-thick BBO (β-barium borate) crystals to generate a spectrum centered around 19360 cm$^{-1}$ and spanning 400 cm$^{-1}$. The average power of the frequency-doubled pulses is 3 mW. The beams are combined on a beam-mixer and interrogate a 90 cm-long iodine cell at room temperature. Their time-domain interference pattern is detected by a 250 MHz bandwidth Si avalanche photodiode, filtered, amplified and digitized by a 16-bit data acquisition board synchronized by an external clock, called adaptive clock [4].

Let us first discuss the need for such an external clock to synchronize the digitization. A well-known consequence of the use of Fourier transformation to sort out the various frequencies of the spectrum is the necessity for sampling the time-domain interferogram within interferometric precision. Otherwise severe artifacts can seriously degrade the spectrum. With Michelson-based Fourier transform spectrometers, such instrumental effects have been studied and successfully minimized for a long time [11]. In dual-comb spectroscopy, new difficulties arise. First, the time intervals between interfering comb pulses can be subject to timing fluctuations that have to be kept lower than 10 attoseconds (relative fluctuations: 10$^{-9}$). Second, the difference of the slippage of the carrier phase relative to the pulse envelope due to laser dispersion between the two combs should be also controlled within 10$^{-9}$. With the digitization of the interferogram at fast sampling rate (e.g. 100 MHz) and the recording time of such interferograms with Doppler-limited resolution on a millisecond scale, very fast servo-control loops would be required together with high accuracy sampling clocks. Successful solutions to circumvent these issues involve stabilizing the combs against continuous-wave lasers with a linewidth of the order of a Hertz and acquiring the interferogram over several seconds [3,8] or monitoring the relative fluctuations of the combs in order to perform computer-based a posteriori corrections [7]. Both approaches however waste the real-time advantage of dual-comb spectroscopy. We have recently developed a simple technique [4] of adaptive dual-comb spectroscopy with free-running lasers that preserves the very short acquisition times and have demonstrated it with 1.5 $\mu$m erbium-doped fiber lasers. Here we extend this technique to frequency-doubled laser systems.

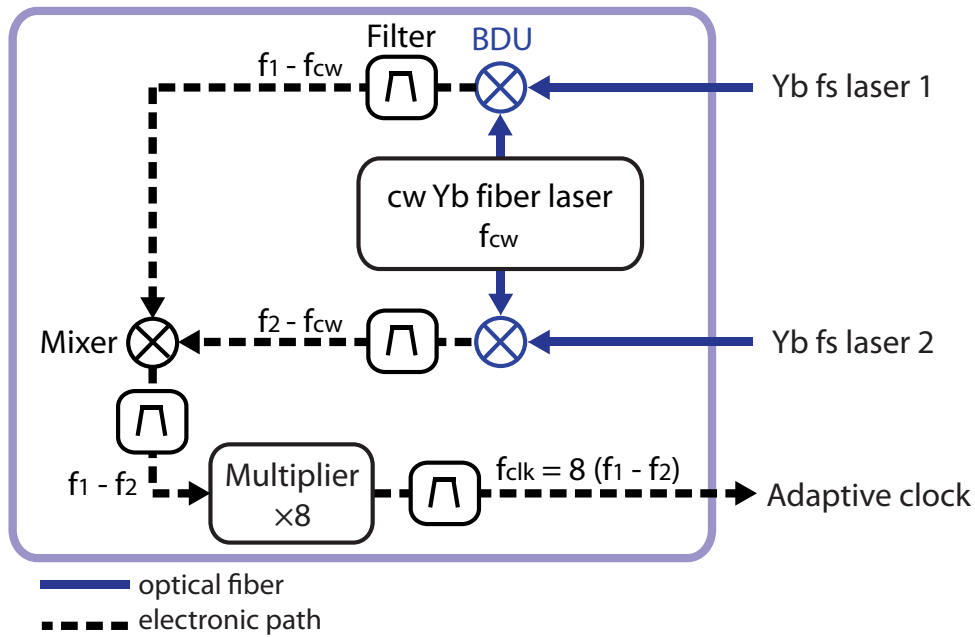

**Figure 2.** Adaptive clock generation scheme. Each fs oscillator beats with a cw free-running Yb fiber laser with a frequency $f_{cw}$ in a beat detection unit (BDU) to isolate a single mode of each fs laser with a frequency of $f_1$ and $f_2$, respectively. The two beating signals are electronically mixed in order to produce an electric signal $f_1-f_2$ reporting the relative fluctuations between the two modes. This filtered and frequency-multiplied signal serves as the external clock to the digitizer.

Figure 2 displays the details of the adaptive clock signal generation, represented by a box in Fig. 1. Each of the two femtosecond ytterbium-doped fiber lasers beats with the same free-running narrow-linewidth continuous-wave (cw) ytterbium-doped fiber laser that emits at 9618 cm$^{-1}$. The beat notes between the femtosecond and cw lasers isolate though proper electronic filtering a single comb line of each comb. The two beating signals are then electronically mixed and the contribution of the cw laser vanishes. A signal at the frequency $f_1-f_2$=10.5 MHz is thus produced. The timing and phase fluctuations between the interfering combs are directly imprinted onto this beating signal. Interference between pairs of optical comb lines image the optical absorption spectrum into the radio frequency region, ideally to cover the full free spectral range which is half of the repetition frequency of the fs lasers i.e. 50 MHz. Therefore, we frequency-multiply eight-fold the 10.5 MHz signal to avoid aliasing. The resulting $f_{clk}$=84 MHz signal provides the adaptive clock signal that triggers the data acquisition. Not shown in the figures, the delays mostly induced by the various electronic components are compensated for. An electronic delay line is therefore inserted after the detector signal in the interferometric spectroscopy set-up and after one of the optical beat detection units in adaptive clock generation device.

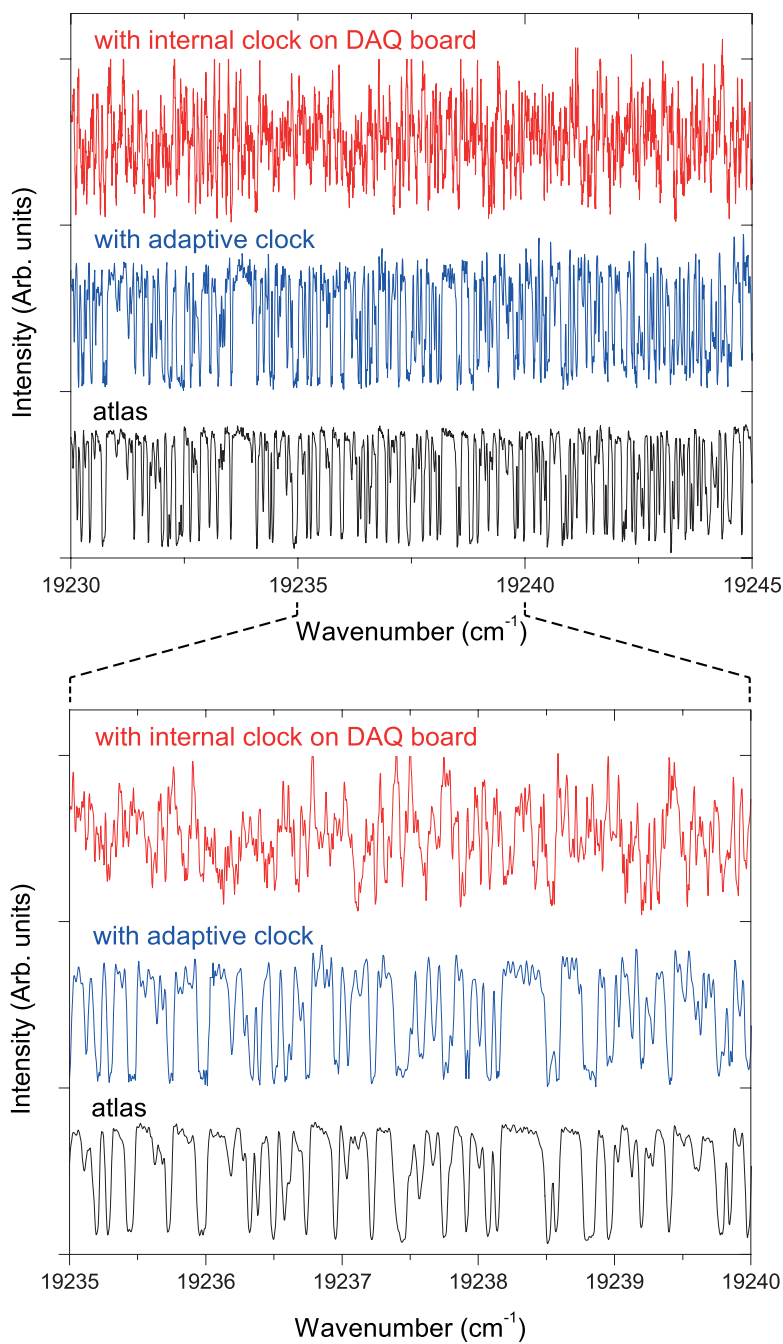

**Figure 3.** Iodine absorption spectra at a resolution of 0.02 cm$^{-1}$ with two different degrees of zoom. For each inset: (upper) dual-comb spectrum with free-running lasers and sampled at the constant clock rate of the digitizer (middle) dual-comb spectrum with free-running lasers and with the adaptive clock, (lower) reproduced from the atlas of Ref. [13]. The dual-comb spectra are measured both within 12 ms in identical experimental conditions but the clock for digitization. The spectrum of [13] is measured with a 50 cm-long I$_2$ cell which explains why the lines in our adaptive spectrum look more intense.

A portion of 15 cm cm$^{-1}$ of the absorption spectrum of iodine around 19240 cm$^{-1}$ is shown in Fig. 3. The full spectrum spans 400 cm$^{-1}$ and is measured within 12 ms, without averaging, at a Doppler-limited resolution of 0.02 cm$^{-1}$ with triangular apodization. The signal to noise ratio is about 30. The spectrum is measured in the region of the center of the strong *B-X* 39-0 band [12]. The adaptive spectrum is compared to a spectrum resulting from an interferogram sampled at the constant clock rate of the data acquisition digitizer, and to the spectrum [13] of the ascii iodine atlas measured by Michelson-based Fourier transform spectroscopy. While the spectrum sampled at the constant clock rate of the data acquisition digitizer is strongly distorted, the adaptive spectrum shows excellent agreement with the reference data from the iodine atlas. The adaptive scheme performed at the fundamental wavelength of the laser therefore proves fully successful even when non-linear frequency conversion is implemented in the spectroscopy set-up. Consistently with our results reported in [4], we find that, even if we compensate for the fluctuations of the femtosecond lasers at a given optical frequency only, the adaptive dual-comb spectra agree with the atlas reference spectrum over 30 cm$^{-1}$ of our spectral span. The spectrum reported in [13] spans 1400 cm$^{-1}$ between 19100 and 20500 cm$^{-1}$ and has been measured within 8 hours at a resolution of 0.02 cm$^{-1}$. While 30 cm$^{-1}$ spectral span is about 50 times narrower, our measurement time is 2 10$^6$ shorter. In high-resolution spectroscopy of atoms and molecules, iodine has been for decades a convenient frequency standard for calibration purposes due to its dense grid in a large part of the visible spectrum. Therefore, besides its interest as an appropriate test for our technique, the possibility to measure the highly-crowded iodine spectrum across 30 cm$^{-1}$ within 12 ms offers an opportunity for frequency calibration of real-time spectroscopic data, as the free-running femtosecond lasers require calibration against molecular lines present in the spectrum.

We have demonstrated a dual-comb spectrometer capable of fully resolving the crowded Doppler-broadened spectrum of iodine in the green region over 30 cm$^{-1}$ of spectral bandwidth within 12 ms. The spectrometer only requires free-running femtosecond laser oscillators and therefore the technique of dual-comb spectroscopy is dramatically simplified. As the adaptive clock generation can successfully compensate for the relative fluctuations of the laser systems when it is performed at the fundamental wavelength of the oscillators, our versatile scheme allows for straightforwardly reaching new spectral territories through nonlinear frequency conversion of near-infrared ultrafast laser sources. This may prove useful of particular relevance in the vacuum and extreme ultra-violet ranges where combs may be generated through high harmonic generation, as continuous-wave lasers are hardly available in these regions.

We warmly acknowledge experimental support by B. Bernhardt. Support by the Max Planck Foundation, the Munich Center for Advanced Photonics, the European Laboratory for Frequency Comb Spectroscopy, Eurostars and the Agence Nationale de la Recherche are acknowledged.